\documentclass[aps,pre,onecolumn]{revtex4}
\usepackage{amssymb}
\usepackage{eurosym}
\usepackage{epsfig}

\begin{document}

\title{Slipping flows and their breaking}
\author{E.A. Kuznetsov $^{(1),(2),(3)}$, E.A. Mikhailov $^{(1),(2),(4)}$}
\affiliation{{\small \textit{$^{(1)}$ - Lebedev Physical Institute, 53 Leninsky
Ave., 119991 Moscow, Russia\\
$^{(2)}$ - Skolkovo Institute of Science and Technology, 1 30 Bolshoy
Boulevard, Skolkovo, 121205 Moscow, Russia;}}\\
{\small \textit{$^{(3)}$ Landau Institute for Theoretical Physics,
Chernogolovka, Moscow region, 142432 Russia; }}\\
{\small \textit{$^{(4)}$ - Moscow State University, 1 Leninskie gori, 119991
Moscow, Russia }}}

\begin{abstract}
The process of breaking of inviscid incompressible flows along a rigid body
with slipping boundary conditions is studied. Such slipping flows are
compressible, which is the main reason for the formation of a singularity
for the gradient of the velocity component parallel to rigid border.
Slipping flows are studied analytically in the framework of two- and
three-dimensional inviscid Prandtl equations. Criteria for a gradient
catastrophe are found in both cases. For 2D Prandtl equations breaking takes
place both for the parallel velocity along the boundary and for the
vorticity gradient. For three-dimensional Prandtl flows, breaking, i.e. the
formation of a fold in a finite time, occurs for the symmetric part of the
velocity gradient tensor, as well as for the antisymmetric part - vorticity.
The problem of the formation of velocity gradients for flows between two
parallel plates is studied numerically in the framework of two-dimensional
Euler equations. It is shown that the maximum velocity gradient grows
exponentially with time on a rigid boundary with a simultaneous increase in
the vorticity gradient according to a double exponential law. Careful
analysis shows that this process is nothing more than the folding, with a
power-law relationship between the maximum velocity gradient and its width: $%
\max|u_x|\propto \ell^{-2/3}$.
\end{abstract}

\maketitle

\section{Introduction}

Collapse as a process of the singularity formation for smooth initial
conditions represents one of the key issues for understanding nature for
hydrodynamic turbulence. The Kolmogorov-Obukhov theory \cite{Kolmogorov,
Obukhov} of developed hydrodynamic turbulence at large Reynolds numbers, ${Re%
}\gg 1,$ in inertial interval predicts the divergence of vorticity
fluctuations $\left\langle \delta \omega \right\rangle $ with scale $\ell $
at small $\ell $ like $\ell ^{-2/3}$, which indicates the connection of
Kolmogorov turbulence with collapse.

Different numerical experiments executed in the late 90s seem to indicate
the observation of collapse, with a more accurate examination which showed its
absence (a discussion of these issues can be found in \cite%
{chae2008incompressible, gibbon2008three}).

Up to now, the question about existence of blow-up for ideal fluids within the
three-dimensional Euler equations remains controversial. It is well-known that in
the two-dimensional Euler hydrodynamics for smooth initial conditions,
collapse as a process of the singularity formation in a finite time, is forbidden
\cite{wolibner, kato, yudovich}. But this, however, does not exclude the
singularity appearance in infinite time with exponential growth, as it
was evidenced in numerical experiments \cite{KNNR}, in which the formation of
the vorticity quasi-shocks is accompanied by exponential in time narrowing  
their widths. 
In the three-dimensional Euler hydrodynamics numerical
experiments also show an exponential increase of the vorticity $\omega $ in
the pancake-type vortex structures for which the exponential narrowing of
the pancake thickness $\ell $ was also observed, but without any tendency to
blow-up \cite{AgafontsevKuznetsovMailybaev2015,
AgafontsevKuznetsovMailybaev2016, AgafontsevKuznetsovMailybaev2017}. The
formation of such structures is possible, as shown in \cite{KuznetsovRuban,
KuznetsovRuban2000, Kuznetsov2002}, for the frozen-in vorticity $\mathbb{%
\omega }$\ within the three-dimensional Euler equations and the vorticity
rotor $\mathbf{B}$ (divorticity) for two-dimensional flows \cite{KNNR}. Due
to this property, frozen-in vector fields turn out to be compressible.
Moreover, it was found that the formation of such structures can be
considered as a folding process when the maximal values of $\omega _{\max }$
and $B_{\max }$ are evaluated proportionally to their widths as $\ell
^{-2/3} $ \cite%
{AgafontsevKuznetsovMailybaev2015,AgafontsevKuznetsovMailybaev2016,AgafontsevKuznetsovMailybaev2017, KuznetsovSereshchenko2019}
(see also review paper \cite{AgafontsevKuznetsovMailybaevSereshchenko2022}
and references therein). The key point for understanding the compressibility
property for frozen-in fluid fields is based on the vortex line
representation introduced for the first time by Kuznetsov and Ruban \cite%
{KuznetsovRuban}. Explanation of compressibility for frozen fields $\mathbf{B%
}$ in incompressible fluids follows from a simple observation. The equation
of motion for $\mathbf{B}$:
\[
\frac{\partial \mathbf{B}}{\partial t}=\mbox{curl}\left[\mathbf{v}\times
\mathbf{B}\right] ,\mbox{ div }\,\mathbf{v=0,}
\]
contains the vector product. Therefore the only velocity component $\mathbf{v%
}_{\perp }$, perpendicular to the field $\mathbf{B}$, can change its value.
The parallel velocity component obviously has no influence on the motion of $%
\mathbf{B}$. In the general case, however $\mbox{div}\, \mathbf{v}_{\perp }%
\mathbf{\neq 0}$. Moreover, due to the frozenness this velocity component
represents a velocity of the field $\mathbf{B}$. This is a reason of the
compressibility of such fields.

It is necessary to mention that all numerical simulations presented in \cite%
{AgafontsevKuznetsovMailybaev2015,AgafontsevKuznetsovMailybaev2016,AgafontsevKuznetsovMailybaev2017,KuznetsovSereshchenko2019,KuznetsovMikhailov2020}
were mainly performed for spatially periodical boxes.

However, recent findings for flows of ideal fluids in the presence of
boundaries, both analytical and numerical, demonstrate blow-up behavior. For
two-dimensional planar flows in the region with non-smooth boundaries
Kiselev and Zlatos \cite{Kiselev} proved blow-up existence.

In 2015 Luo and Hou \cite{Hou} in their intriguing numerical experiments for
axi-symmetrical Eulerian flows with swirl inside the cylinder of constant
radius observed appearance of collapse just on the boundary. It was a
challenge why boundaries play so important role in formation of
singularities.

In 2019 Elgindi and Jeong \cite{elgindi2019finite} proved the existence of
finite-energy strong solutions to the axi-symmetric 3D Euler equations
\textit{outside} the cylinder $(1 + \epsilon |z|)^2 \leq x^2+y^2$ which
become singular in finite time. Singularity develops on the cylinder surface
at $z=0, x^2+y^2=1$ corresponding to a corner of this cylinder. The flow
geometry, thus, dictates appearance of the singularity point.The latter
result correlates with studies of Kiselev and Zlatos \cite{Kiselev}) for
two-dimensional Euler flow inside the region with not-smooth boundaries.

The similar result was obtained recently in \cite{elgindi2020finite} by
Elgindi and Jeong for finite-time singularity formation for the 2D
Boussinesq system. The flow region again was not smooth, contains a corner
and singularity develops again at the surface corner. The authors of \cite%
{elgindi2020finite} state that the flow region can be prolonged up to the
half-space. However, from our point of view, just the flow geometry in both
cases \cite{elgindi2019finite} and \cite{elgindi2020finite} dictates
appearance of the singularity point but in less extent than the initial
conditions. For smooth boundary conditions in the case of two-dimensional
Euler equations for flows inside a disk Kiselev and {\v{S}}ver{\'{a}}k \cite%
{kiselev2014small} constructed an initial data for which the gradient of
vorticity exhibits double exponential growth in time with maximum value on
the boundary. Simultaneously the velocity gradient grows on the boundary
exponentially in time.

In this paper we show that flat boundary itself introduces some element of
compressibility into flow which, from our point of view, can be considered
as a reason of the singularity formation on the boundary. We will consider
the 2D and 3D inviscid Prandtl equations which describes the dynamics of the
boundary layer, and demonstrate that singularity is formed for the velocity
gradient on the wall with slipping boundary conditions. This process is
nothing more than breaking phenomenon which is well known in gas dynamics
since the classical works of famous Riemann. Notice that in 1985 E and
Engquist \cite{E-Engquist} reported some rigorous results about blow-up
existence for both inviscid and viscous Prandtl equations for some initial
data when the velocity component parallel to a wall vanishes at the whole
vertical line. For such initial conditions these authors found sufficient
condition for blow-up in the viscous case and exact blow-up solution for the
inviscid Prandtl equation depending on this vertical coordinate. It is worth
noting that before, in 1980, the blowup appearance in the Prandtl equations
was observed in the numerical simulations by Van Dommelen and Shen \cite%
{DS80} (see also the review \cite{kukavica2017van} and references
therein).

Following to \cite{kiselev2014small}, in this paper on the example of flows
between two parallel walls we will show numerically and give some analytical
arguments that double exponential growth for vorticity gradient and
exponential growth for velocity gradient at the boundary, are connected each
others. From our point of view, explanation of the results by Lou and Hou
\cite{Hou} as well as by Kiselev and {\v{S}}ver{\'{a}}k \cite%
{kiselev2014small} are connected with slipping flows for smooth boundaries.
For such type of flows the normal velocity component vanishes on the
boundary and the rest slipping flow along the boundary will be considered as
a compressible one. The divergence of the normal velocity, in this
situation, provides incompressibility condition of the flow. In the case of
the two-dimensional inviscid Prandtl equation, first time this fact was
established by Hong and Hunter \cite{HongHunter} for the pressureless
conditions. In this paper we show that for the constant pressure gradient it
is also possible to find breaking criteria and establish blow-up
for the vorticity gradient on the boundary. The latter, as we will show,
occurs in the correspondence with double exponential growth of the vorticity
gradient for the 2D Euler flows between two parallel plates. We show that
the maximum of the velocity gradient $\max |u_{x}|$ at the wall grows in
time exponentially like for a disk \cite{kiselev2014small}. This process can
be considered as folding with typical dependence between growing $\max
|u_{x}|$ and its narrowing in time width $\ell $:
\[
\max |u_{x}|\propto \ell ^{-2/3}.
\]
For the 3D Prandtl equations the slipping flow in the pressureless case is
defined from the 2D Hopf equation for two velocity components parallel to
the wall. The gradient catastrophe here is of the blow-up type for both
stress velocity tensor and vorticity.

The plan of the paper is as follows. In the next section we consider 2D
ideal flows in the boundary layer and analyze them by means of the so-called Crocco
transformation \cite{crocco1941sullo} which we slightly modify. It is worth
noting that the Crocco transformation can be applied in more general cases,
for instance, when pressure depends on coordinate. Note also that the Crocco
transformation was widely used by Oleinik \cite{oleinik}.

Based on the Crocco transformation, we show that the 2D inviscid Prandtl
equations for incompressible flow with constant pressure can be transformed
into the linear relation between velocity component $u$ parallel to the wall
and streamfunction $\psi $. This linear relation can be resolved by
introducing the generating function $\theta $. For the Prandtl equations $%
\theta $ is easily found and the general solution can be obtained in an
implicit form for the slipping boundary condition. In this case, the
singularity of the gradient type develops on the wall due to the
compressible character of the slipping flow. The similar character for the
singularity formation remains for arbitrary dependence of pressure on the
coordinate $x$ along the boundary. The velocity gradient near singular point
behaves like $(t_{0}-t)^{-1}$ where $t_{0}$ is the collapse time.

In the third section we present results of numerical simulations for flows
within 2D Euler equation between two parallel plates. Careful analysis of
the numerical results allows us to say that the growth of the velocity
gradient at the boundary can be considered as a folding process.

Section 4 deals with flows in the framework of the 3D inviscid Prandtl
equation. In this case, the slipping flow in the pressureless limit is
defined from the 2D Hopf equation for two velocity components parallel to
the wall. The gradient catastrophe here is of the blow-up type for both
stress velocity tensor and vorticity. The last Section is the conclusion.

We also note that the preliminary results of this paper about slipping flows
within the inviscid Prandtl equations were presented by Kuznetsov at the
International conference "Integrability", September 22 - 24, 2021 \cite%
{kuznetsov2021}.

\section{Two-dimensional Prandtl equations}

In the inviscid limit the Prandtl equation for 2D flows is written for the
velocity component parallel to the blowing plane $y=0$:
\begin{equation}
u_{t}+uu_{x}+vu_{y}=-P_{x},\,\,u_{x}+v_{y}=0  \label{prandtl}
\end{equation}%
with the following initial-boundary conditions:
\begin{equation}
u(x,y,0) =u_{0}(x,y),\,\,v(x,y,0)=v_{0}(x,y),  \label{initial-u}
\end{equation}
\[
\mbox{and}\,\,v|_{y=0} =0,\,\,\lim_{y\rightarrow \infty }u(x,y,t)=U(x),
\]
where $u$ and $v$ are $x$- and $y$-velocity components, respectively.
Subscripts here and everywhere below mean the corresponding derivatives.

The Prandtl approximation assumes that the boundary layer has a thickness
much smaller than the characterized size along the layer. As a
result pressure $P$ can be considered as a function depending on the
longitudinal coordinate $x$ only.

Introducing stream function $\psi $,
\begin{equation}
u=\psi _{y},\,\,v=-\psi _{x},  \label{stream}
\end{equation}%
the initial-boundary conditions (\ref{initial-u}) are rewritten as
\[
\psi (x,y,0)=\psi _{0}(x,y),\psi (x,0,t)=\mathrm{const}(=0),\,\psi
(x,y,t)\rightarrow U(x)y\mbox{ at }y\rightarrow \infty .
\]
Here we assume that the pressure $P$ being independent on $y$ and $%
t $ satisfies the Bernoulli law at $y\rightarrow \infty $
\[
\frac{U^{2}(x)}{2}+P(x)=\mbox{const}.
\]
The inviscid Prandtl equation (\ref{prandtl}) can be used for approximate
description of real flows in the boundary layer if one assumes its thickness
to be large enough in comparison with a width of the viscous sublayer. In
the latter case the influence of viscous sublayer might be modeled by the
slip boundary condition, $v(x,0)=0$. In this paper for the equation (\ref%
{prandtl}) we will use only slip boundary conditions on the wall.

It is worth noting that within the Prandtl approximation for inviscid flows
it is possible to introduce the vorticity as
\[
\omega =-\frac{\partial u}{\partial y}
\]
which satisfies the equation of the same form as for the 2D Euler fluids:%
\begin{equation}
\omega _{t}+u\omega _{x}+v\omega _{y}=0.  \label{vorticity}
\end{equation}%
Thus, $\omega $ is the Lagrangian invariant. By this reason, its values will
be bounded at all $t>0$. However, for another components of the velocity gradient
such restrictions are absent. As we will see below, $u_{x}$ as well as $%
v_{y} $ can take arbitrary values, in particular, infinite ones.

Firstly we consider the special case $P=\mathrm{const}$ when the quantity $u$
in (\ref{prandtl}) becomes a Lagrangian invariant as well as, for instance,
the density $n$ in the Boussinesq system.

Consider the equation for $n$ advected by the fluid,
\begin{equation}
n_{t}+un_{x}+vn_{y}=0  \label{lagr}
\end{equation}%
with condition
\[
u_{x}+v_{y}=0.
\]
Suppose that we know solution of equation (\ref{lagr})
\begin{equation}
n=n(x,y,t).  \label{sol1}
\end{equation}%
Assuming this function is smooth and monotonic at each $x$ with increase (or
decrease) of $y$. In this case one can find
\begin{equation}
y=y(x,n,t)  \label{map}
\end{equation}%
inverse to (\ref{sol1}). Thus, at fixed $t$ this function defines a mapping
from old ($x,y$) variables to a new independent Lagrangian variable $n$ and
old Eulerian coordinate $x$ so that we arrive at the mixed
Lagrangian-Eulerian description which can be considered as a \textit{%
non-complete} Legendre transformation. In gas dynamics, as well known the
classical example of a \textit{complete} Legendre transformation represents
the Hodograph transformation when instead of density $n(x,t)$ and velocity $%
v(x,t)$ the inverse functions $t(n,v)$ and $x(n,v)$ are considered.

Fixing $n$ in (\ref{sol1}, \ref{map}) yields the $n$-level line. Thus, this
transformation represents transition to the \textit{curvilinear} system of
coordinates movable with level lines of $n$. In terms of the new Lagrangian
variable $n$ and the old ones $x$ and $t$, as we see now, the equation (\ref%
{lagr}) transforms into a linear equation.

To rewrite the equation in new variables one can find how derivatives with
respect to variables $(x,y,t)$ and derivatives relative to $(x,n,t)$ are
connected with each other:
\begin{eqnarray}
\frac{\partial f}{\partial t} &=&\frac{1}{y_{n}}\,[f_{t}y_{n}-f_{n}y_{t}],
\label{der1} \\
\frac{\partial f}{\partial x} &=&\frac{1}{y_{n}}\,[f_{x}y_{n}-f_{n}y_{x}],
\label{der2} \\
\frac{\partial f}{\partial y} &=&\frac{f_{n}}{y_{n}}.  \label{der3}
\end{eqnarray}%
In these relations derivatives standing in the left hand sides are taken
against $(x,y,t)$, and, respectively, in the right hand sides relative to $%
(x,n,t)$. The expression $y_{n}$ standing in denominator in these relations
is nothing more than the Jacobian of mapping (\ref{map}). This Jacobian is
not fixed, for instance, it is not equal to unity, because we use the mixed
Lagrangian-Eulerian description. Our change of variables given by (\ref{map}%
) will be correct up to the moment of time when the Jacobian vanishes: $%
J\rightarrow 0$. Generally, it will happen first time in one separate point.
If initial data have some symmetry then the situation may be more
degenerated. The Jacobian in such situation can vanish simultaneously in a
few points or even at some curve, finite or infinite.

Applying the formulas (\ref{der1}), (\ref{der3}) to the equation (\ref{lagr})
gives us the kinematic condition, well-known for the free-surface
hydrodynamics:
\begin{equation}
y_{t}+uy_{x}=v.  \label{y}
\end{equation}%
Expressing the velocity through the streamfunction $\psi $ (\ref{stream}) in
terms of ($x,n$) by means of relations (\ref{der1}-\ref{der3}) gives
\begin{equation}
u=\frac{1}{y_{n}}\,\psi _{n},\,\,v=-\psi _{x}+\frac{y_{x}}{y_{n}}\,\psi _{n}.
\label{u-v}
\end{equation}%
Substitution of these formulas into the equation (\ref{y}) results in the
linear relation between $y$ and $\psi $:
\[
y_{t}=-\psi _{x}.
\]
Note, that in this equation all derivatives are taken for fixed $n$. This
equation can be easily resolved by introducing the generating function $%
\theta (x,n,t)$:
\begin{equation}
y=\theta _{x},\,\,\psi =-\theta _{t}.  \label{theta}
\end{equation}%
To find the function $\theta (x,n,t)$ one needs to know dynamical behavior
of the velocity field.

\section{Solution to the 2D inviscid Prandtl equations}

First consider the case of zeroth pressure gradient, $P_{x}=0$. In this
case, to find $\theta $ for the inviscid Prandtl equation ({\ref{prandtl}})
we need to solve the first equation of the system (\ref{u-v}) (by changing $%
n $ to $u$):
\begin{equation}
u=\frac{1}{y_{u}}\,\psi _{u},  \label{main}
\end{equation}%
which after substitution of (\ref{theta}) transforms into the linear
equation
\[
\frac{\partial \theta _{u}}{\partial t}+u\,\frac{\partial \theta _{u}}{%
\partial x}=0.
\]
This equation evidently has the following solution:
\[
\theta _{u}=F(x-ut,u)
\]
where $F$ is an arbitrary smooth function which should be determined from
the boundary-initial conditions. Integration with respect to $u$ yields
\[
\theta =\int_{f(x,t)}^{u}F(x-zt,z)dz+g(x,t)
\]
where $f(x,t)$ and $g(x,t)$ are arbitrary functions of two arguments. Using
the definition of the generating function (\ref{theta}) we have
\[
y =\int_{f(x,t)}^{u}\frac{\partial }{\partial x}F[x-zt,z]dz+g_{x}(x,t)-F%
\left[ x-f(x,t)t,f(x,t)\right] f_{x}(x,t),
\]
\[
\psi =-\int_{f(x,t)}^{u}\frac{\partial }{\partial t}%
F[x-zt,z]dz-g_{t}(x,t)+F[x-f(x,t)t,f(x,t)]f_{t}(x,t).
\]
Here both functions $f(x,t)$ and $g(x,t)$ should be determined from the
boundary conditions. It turns out that the function $f(x,t)$ coincides with $%
u$ at $y=0$ and $g(x,t)=0$ so that
\[
y =\int_{f(x,t)}^{u}\frac{\partial }{\partial x}F[x-zt,z]dz
\]
\[
\psi =-\int_{f(x,t)}^{u}\frac{\partial }{\partial t}F[x-zt,z]dz.
\]
The formulas (\ref{theta}) and (\ref{main}) coincide with  the Crocco transformation \cite{crocco1941sullo} 
which was  applied first time to the 2D Prandtl equation but written here in the different form.

\subsection{The Hopf equation}

Because at $y=0$ (at the wall) the normal velocity component $v$ vanishes,
the Prandtl equation (\ref{prandtl}) at $P_{x}=0$ becomes nothing more than
the Hopf equation (i.e., this is an autonomous equation!),%
\begin{equation}
u_{t}+uu_{x}=0.  \label{Hopf-equation}
\end{equation}%
This equation is well-known and has been discussed many times in the
literature (see, for instance \cite{Frisch}). This is the simplest equation
describing breaking, in particular, in a one-dimensional gas flow with zero
pressure. We briefly show how the corresponding solution of this type can be obtained.

It is well known that solution to (\ref{Hopf-equation}) is written in the following implicit form%
\[
u=u_{0}(a),
\]%
\[
x=a+u_{0}(a)t
\]
or%
\begin{equation}
u=u_{0}(x-ut).
\end{equation}%
This means that on the boundary the breaking, i.e. the formation of
singularity in a finite time, becomes possible. It happens when the
derivative
\[
\frac{\partial u}{\partial x}=\frac{u_{0}^{\prime }(a)}{1+u_{0}^{\prime }(a)t%
}=\frac{1}{t+(u_{0}^{\prime }(a))^{-1}}
\]
at some point $x=x_{\ast }$ first time, $t=t_{\ast }$, becomes infinite. It
is evident that in this case
\begin{equation}
t_{\ast }=\min_{a}\left[ -1/u_{0}^{\prime }(a)\right] >0.  \label{minimum}
\end{equation}%
If, for instance, $u_{0}(a)=A\cos a$ with $a\in \lbrack 0,\pi ]$ the
singular point $a_{\ast }=\pi /2$ and $t_{\ast }=1/A$. Near the breaking
point the derivative $\frac{\partial u}{\partial x}$ behaves like%
\begin{equation}
\frac{\partial u}{\partial x}\simeq -\frac{1}{\tau +\beta (\Delta a)^{2}}
\label{breaking-point}
\end{equation}%
where
\[
\tau =t_{\ast }-t,\,\,\Delta a=a-a_{\ast },
\]
\[
\beta =1/2\frac{d^{2}}{da^{2}}\left[ 1/u_{0}^{\prime }(a)\right]
|_{a=a_{\ast }}.
\]%
Thus, asymptotically this dependence demonstrates a self-similar
compression, $\Delta a\propto \tau ^{1/2}$. Notice, the denominator in (\ref%
{breaking-point}), up to the constant multiplier $C$, coincides with the
Jacobian,
\begin{equation}
J=\frac{\partial x}{\partial a}=C\left( \tau +\beta a^{2}\right).
\label{derivative}
\end{equation}%
where for simplicity we put $a_{\ast }=0$. Integration of this equation
gives the cubic dependence relative to $a$,%
\[
x=C\left( \tau a+\beta a^{3}/3\right) .
\]
Respectively, in the physical space we get more rapid compression than in
the space of the Lagrangian markers: $x\propto \tau ^{3/2}$.

At distances $\beta a^{2}\gg \tau $, for the Jacobian we have the
time-independent asymptotics,
\[
J\sim {x^{2/3}}.
\]
Thus, as $\tau \rightarrow 0$ we arrive at the singularity for the
derivative $u_{x}$,
\begin{equation}
u_{x}\propto x^{-2/3}.  \label{1Dsingularity}
\end{equation}%
Any time changes of $u_{x}$ happen at the narrowing region in the $a$-space,
$a\propto \tau ^{1/2}$, or equivalently at $x\propto \tau ^{3/2}$. In the
physical space, thus, we have the following self-similar asymptotics,
\begin{equation}
u_{x}\simeq \frac{1}{\tau }F(\xi ),\,\,\xi =\frac{x}{\tau ^{3/2}}
\label{similarity}
\end{equation}%
where function $F(\xi )$ at large $\xi $ behaves proportionally to $\xi
^{-2/3}$. Hence the maximum value of $u_{x}$ and its width $\ell $ \ are
connected each other by the power-type law:
\[
\max u_{x}\propto \ell ^{-2/3}.
\]
It should be emphasized that this is a general asymptotic behavior for
folding, independently whether the singularity formation happens in finite
or infinite time.

It is interesting to note that for the constant pressure gradient, $P_{x}=%
\mathrm{const}$, the equation for slipping flow (at $y=0$)
\begin{equation}
u_{t}+uu_{x}=-P_{x}  \label{const}
\end{equation}%
can be solved by the same means as (\ref{Hopf-equation}) (compare with \cite%
{FrolovskayaPukhnachev}):%
\[
u =u_{0}(a)-P_{x}t,
\]
\[
x =a+u_{0}(a)t-P_{x}t^{2}/2.
\]
Breaking moment of time is given formally by the same formula (\ref{minimum}%
),%
\[
\frac{\partial u}{\partial x}=\frac{u_{0}^{\prime }(a)}{1+u_{0}^{\prime }(a)t%
},
\]
\[
t_{\ast }=\min_{a}\left[ -1/u_{0}^{\prime }(a)\right] .
\]
For arbitrary dependence $P(x)$ a solution of (\ref{const}) is found by
means of the method of characteristics. The equations for characteristics
\[
\frac{d}{dt}u=-P_{x},\frac{d}{dt}x=u.
\]
reduce to the Newton equation for $x(t)$:
\[
\frac{d^{2}x}{dt^{2}}=-P_{x}.
\]
The first integral is the energy,%
\[
E(a)=\frac{{x_t}^{2}}{2}+P(x)=\frac{u_{0}^{2}(a)}{2}+P(a),
\]
that allows one to integrate the equation in quadratures%
\[
\int_{a}^{x}\frac{dz}{\sqrt{2(E-P(z))}}=t,
\]
that defines the mapping $x=x(a,t).$ In this case, the breaking time $%
t_{\ast }$ is found as a minimal root $T(>0)$ for equation $J(a,T)=0,$
\[
t_{\ast }=\min_{a}T(a),
\]
where $J=\partial x/\partial a$ is a Jacobian of the mapping.

The singularity for the velocity gradient on the boundary, which we obtain,
appears as a result of collision of two counter-propagating slipping
flows. As it was demonstrated numerically first time by Dommelen and Shen
\cite{DS80} and later confirmed by Hong and Hunter \cite{HongHunter} this
interaction leads to the formation of jets in perpendicular to the boundary
direction (see also \cite{kukavica2017van}). As it will be shown in the next
section such behavior remains for flows in the 2D Euler equations with
slipping boundaries, where, unlike the 2D Prandtl flows, the maximal
velocity gradient grows in time exponentially on the boundary and this
growth is accompanied also by the generation of jets in perpendicular
direction to the boundary.

\subsection{Behavior for the vorticity gradients on the boundary}

As we saw previously, the Hopf equation (\ref{Hopf-equation}) has a blow-up
behavior if initially $u_{0}^{\prime }(a)<0$. In this case, at the singular
point $a=a^{\ast }$ maximum of the parallel velocity gradient behaves like $%
u_{x}\sim (t_{\ast }-t)^{-1}.$

Now we will calculate how the vorticity $\omega $ behaves at this point. In the
Prandtl approximation, as we mentioned already, the vorticity $\omega
=-u_{y} $ satisfies the equation (\ref{vorticity}).

We are going to show that the vorticity gradient at the maximal point of the velocity
gradient at $y=0$ has the same behavior as the velocity gradient $u_{x}$ .
It is important to remind that the vorticity being the Lagrangian invariant cannot
change its value along the trajectories, but the gradient of $\omega $ can
blow up on the wall.

For simplicity let us restrict by considering the pressureless case. Firstly, we
differentiate (\ref{vorticity}) with respect to $x$ and then in the result
put $y=0$ where $v=0$ and respectively $v_{x}=0$. This yields the following
equation%
\begin{equation}
\frac{\partial \omega _{x}}{\partial t}+u\frac{\partial }{\partial x}\omega
_{x}=-u_{x}\omega _{x}.  \label{divorticity}
\end{equation}%
Secondly, apply to this equation the characteristics method that gives:%
\begin{eqnarray}
\frac{dx}{dt} &=&u(x,t),  \label{x} \\
\frac{d\omega _{x}}{dt} &=&-u_{x}\omega _{x}.
\end{eqnarray}%
It is easy to see that integration of the second equation of this system at
the breaking point where in accordance with (\ref{breaking-point}) $u_{x}$
can be approximated as
\[
u_{x}\simeq \frac{1}{t-t_{\ast }}
\]
gives at this point the same singular behavior for $\omega _{x}$:
\[
\omega _{x}\simeq \frac{A}{t-t_{\ast }},
\]
where $A$ is a constant.

At the end of this section we would like to pay attention to the following
important fact. Folding development for slipping flows always results in the
formation of a jet in the normal direction to the boundary. As it was mentioned
in \cite{HongHunter}, breaking (as a folding happening in a finite time) for
the slipping flows in the 2D Prandtl equation becomes possible because the
pressure gradient normal to the boundary can not prevent the formation of
jets. As we will see in the next section for the 2D Euler equations the
pressure gradient prevents the singularity formation in a finite time and
the velocity gradient growth on the boundary occurs more slowly.

\section{Folding of the slipping flows for the 2D Euler equation}

In this section we consider slipping flows within the 2D Euler equation
between two parallel plates. We present results of numerical simulations for
folding of such flows which develops on the plate boundary. This part of
work was motivated by the paper by Kiselev and {\v{S}}ver{\'{a}}k \cite%
{kiselev2014small} where for the 2D Euler equations the authors constructed
an initial data for which the gradient of vorticity exhibits double
exponential growth in time with maximum value on the boundary. Our numerical
results are in the correspondence with this paper. In particular, we have
observed that the maximum of the velocity gradient $\max |u_{x}|$ at the
wall grows in time approximately exponentially like for a disk. This results
in the double exponential growth of the vorticity gradient for the 2D Euler
flows. We have established also that this process can be considered as
folding with typical dependence between growing $\max |u_{x}|$ and its
narrowing in time width $\ell $: $\max |u_{x}|\propto \ell ^{-2/3}.$

Numerically we solve the 2D Euler equation for the vorticity $\omega $
\begin{equation}
\omega _{t}+u\omega _{x}+v\omega _{y}=0  \label{euler}
\end{equation}%
for flows between two rigid plates $y=0$ and $y=h$, with slip boundary
conditions on the both plates,
\[
v(x,0)=v(x,h)=0,
\]
and $2\pi $-periodical conditions relative to $x$: $[-\pi \leq x\leq \pi ]$.
The velocity components $u$ and $v$ are found through the streamfunction $%
\psi $ by means of standard formulas (\ref{stream}) where the vorticity is
expressed as
\begin{equation}
\omega =-\Delta \psi =-\left( \frac{\partial ^{2}\psi }{\partial x^{2}}+%
\frac{\partial ^{2}\psi }{\partial y^{2}}\right) .  \label{eq2}
\end{equation}%
For the streamfunction $\psi $ we use the zero boundary conditions at $y=0$
and $y=h$. Such choice corresponds to the absence of the flow with a
constant velocity along $x$-direction.

For integration of equations (\ref{euler}) and (\ref{eq2}) we used two
approaches. These are the so-called time marching method (pseudo-transient
method) \cite{Kalitkin} to inverse the Laplace operator and the
Peaceman-Rachford scheme \cite{Peaceman-Rachford} (see also \cite%
{Douglas-Rachford}) for (\ref{euler}). The accuracy for the first method was
$||\Delta \psi +\omega ||^{2} \le 10^{-7}$. The kinetic energy $E=\int dxdy
(u^2/2+v^2/2)$ in our simulations was conserved not worth than $10^{-6}$.

Below we present results of our simulations for the following initial
conditions (IC):
\begin{equation}
\psi (x,y,0)=-By(y-h)^{2}\sin x;  \label{IC}
\end{equation}%
where $h=2$ and $B=0.1$. These IC were chosen so that the folding point
appears at $x=0$ on the  boundary $y=0$.

At first, the spatiotemporal dependencies of velocity were found numerically
and then the temporal evolution of its gradient was determined. Analysis of
the distribution of the velocity gradient showed that for the IC (\ref{IC})
almost from the very beginning its maximum is concentrated on the boundary $%
y=0$ in the vicinity of the point $x=0$ which corresponds to the folding
point. Around this point the parallel velocity $u$ behaves almost like for
overturning describing by the Hopf equation (\ref{Hopf-equation}).
\begin{figure}[th]
\centering
\includegraphics[width=0.7\textwidth]{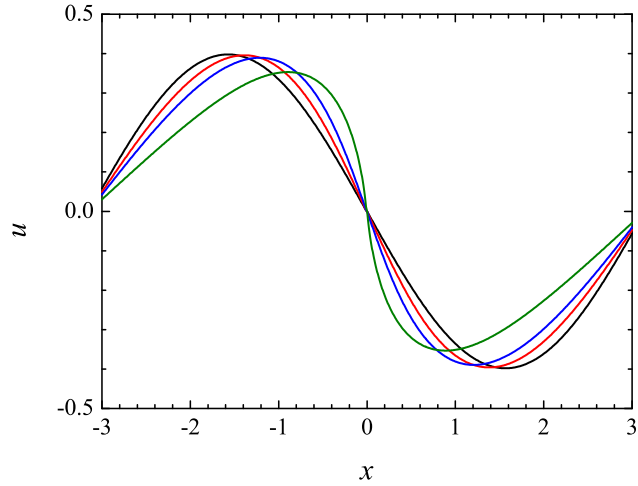}
\caption{Dependencies of the slipping ($y=0$) velocity as a function of $x$
at different moments of time. Black line corresponds to $t=0$, red line -- $%
t=1$, blue line -- $t=2$, green line -- $t=5$.}
\end{figure}
In Fig.1 one can see that at the initial moment the maximal
velocity $u$ increases a little but then becomes more or less constant. The
latter behavior is familiar to the breaking process in the 2D Prandtl
equations for slipping flow within the Hopf equation.
\begin{figure}[th]
\centering
\includegraphics[width=0.7\textwidth]{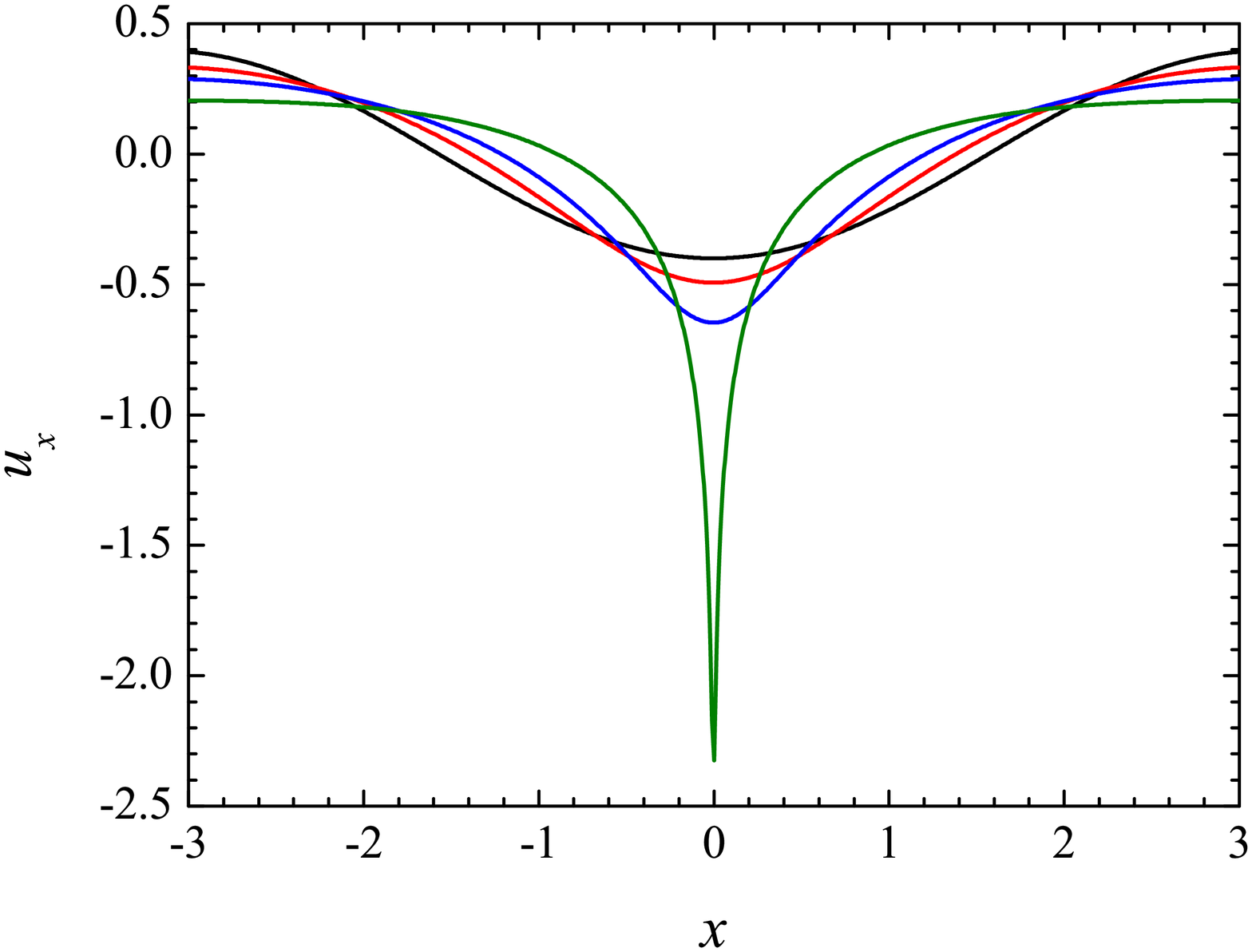}
\caption{Dependencies of the slipping ($y=0$) velocity gradient $u_{x}$ as a
function of $x$ at different moments of time: $t=0$ (black line), $t=1$
(red) , $t=2$ (blue) , $t=5$ (green). With time increasing the profile $u_{x}
$ is seen to transform into a cusp.}
\end{figure}
Fig.2 shows dependencies for the slipping velocity gradient $u_{x}$ as a
function of $x$ at different moments of time. These distributions transform
with time into a cusp with the self-similar dependence $u_{x}$ familiar to
that for the breaking for the slipping flow in the Prandtl approximation (%
\ref{similarity}). Note that $u_{x}$ becomes more and more negative.
\begin{figure}[th]
\centering
\includegraphics[width=0.7\textwidth]{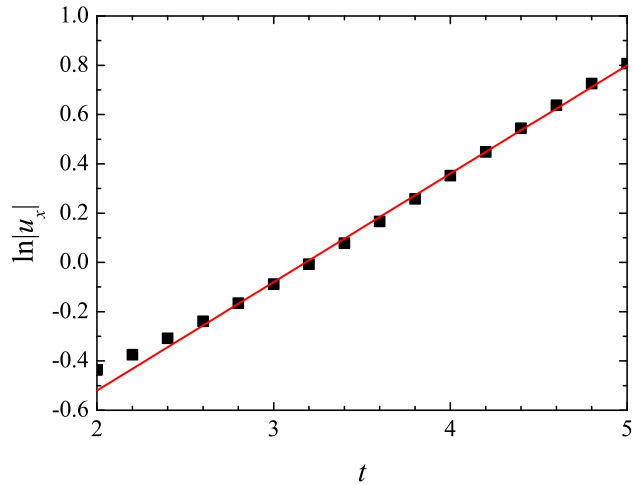}
\caption{Time dependence of the maximal value of $|u_{x}|$ for the slipping (%
$y=0$) flow (logarithmic scale). The black squares correspond to the
numerical results, the red line indicates the slope $\propto e^{\protect%
\gamma _{1}t}$ with $\protect\gamma _{1}=0.44$.}
\end{figure}

\begin{figure}[ht]
\centering
\includegraphics[width=0.7\textwidth]{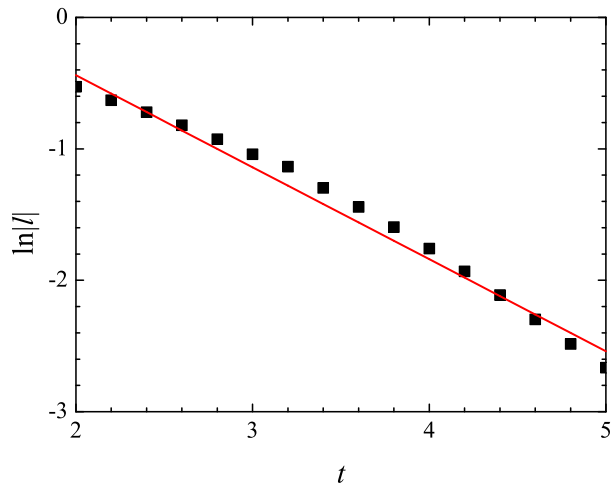}
\caption{Spatial thickness of $|u_x|$ for the slipping ($y=0$) flow as a
function of time. The black squares correspond to the numerical
results, the red line indicates the slope $\propto e^{-\protect\gamma_2t}$
with $\protect\gamma_2 =0.7$.}
\end{figure}

Fig.3 and Fig.4 demonstrate that the maximum $|u_{x}| $ grows approximately
exponentially in time while its thickness $\ell $ decreases but also
exponentially.
\begin{figure}[ht]
\centering
\includegraphics[width=0.7\textwidth]{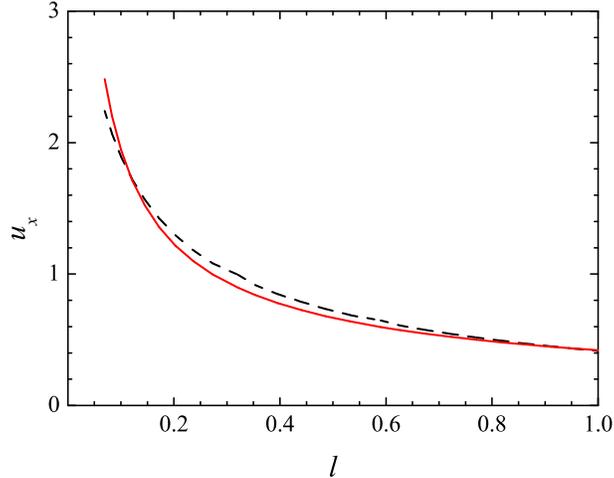}
\caption{Maximum velocity gradient $|u_x|$ versus thickness $\ell$. Black
dotted line corresponds to numerical results, and the red line is the power
dependence $\max|u_x|\propto \ell^{-2/3}$.}
\end{figure}

Such behavior means that as is seen from Fig.5 between $\max |u_{x}|$ and $%
\ell $ the power law dependence arises,
\[
\max |u_{x}|\propto \ell ^{-\alpha },
\]
with exponent $\alpha \approx 2/3$.

This process for the slipping flow can be considered as a folding, it
results in the formation of jet in transverse direction to the boundary $y=0$
(see Fig.5).
\begin{figure}[ht]
\centering
\includegraphics[width=0.7\textwidth]{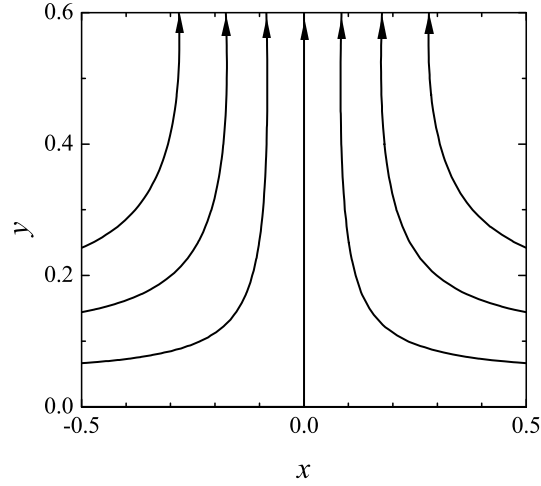}
\caption{Streamlines at $t=5$ in the neighborhood of $x=y=0$ show the jet
forming as a result of folding of the slipping flow.}
\end{figure}
The physical reason of the jet appearance is connected with collision of two
counter-propagating slipping flows. Unlike the Prandtl case this process for
the 2D Euler equation occurs more slowly.

Following arguments to Subsection 3.2 exponential growth of $|u_{x}|$ should
promote the vorticity gradient increase during the folding process. Indeed,
it is so, the 2D Euler equation for $\omega_{x}$ coincides in its form with (%
\ref{divorticity}):
\begin{equation}  \label{divort}
\frac{\partial \omega _{x}}{\partial t}+u\frac{\partial }{\partial x}\omega
_{x}=-u_{x}\omega _{x}.
\end{equation}
It is worth noting that this equation is $y$-component of the equation for
divorticty \cite{KNNR} taken at $y=0$. As before, equation (\ref{divort})
can be solved by the method of characteristics:
\[
\frac{dx}{dt} =u(x,t),
\]
\[
\frac{d\omega _{x}}{dt} =-u_{x}\omega_{x}.
\]
From the second equation of this system we get that
\[
\log \omega _{x}=-\int^{t}u_{x}dt^{\prime }.
\]
As already noted, at the folding region $u_{x}<0$ . If $\max |u_{x}|$
increases exponentially in time for the slipping flow then $\omega _{x}$
will have a double exponential growth in time.
\begin{figure}[ht]
\centering
\includegraphics[width=0.7\textwidth]{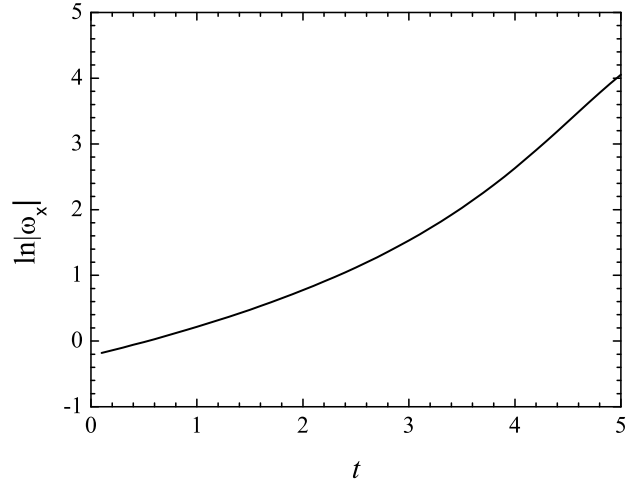}
\caption{Dependence of $\protect\omega _{x}$ at $x=0$ (logarithmic scale)
versus time.}
\end{figure}
Our numerical simulations support this conclusion. In the logarithmic scales
as it is seen from Fig.6 initially the growth of $\ln\omega _{x}$ at the
folding point $x=0$ looks like exponential (direct line) but at the later
stage one can see positive deviation from this line and respectively the
beginning of the double exponential increase of $\omega _{x}$. Thus, our
numerical results correspond to those by Kiselev and {\v{S}}ver{\'{a}}k \cite%
{kiselev2014small} for the Eulerian flow inside a disk.

\section{Breaking in the 3D Prandtl equations}

In this section we consider the 3D inviscid Prandtl equations for which as
it will be seen below the slipping flow introduce more possibilities for
finite-time singularity formation in comparison with the 2D equations. The
3D inviscid Prandtl equations represent a simple generalization of the 2D
Prandtl equations:
\begin{equation}
\mathbf{u}_{t}+(\mathbf{u\cdot }\nabla )\mathbf{u}+v\mathbf{u}_{z}=-\nabla P(%
\mathbf{r}),\,\,(\nabla \cdot \mathbf{u})+v_{z}=0.  \label{3DP}
\end{equation}%
Here $\mathbf{r}=(x,y)$ and $\mathbf{u}$ are coordinates and velocity
components parallel to the wall, respectively, $\nabla =(\partial
_{x},\partial _{y})$, $v$ is the normal ($\Vert \widehat{z}$) velocity
component. Boundary conditions to these equations at the wall are slipping
ones:
\[
v|_{z=0}=0,
\]
another boundary condition at $z\to \infty$ connects the asymptotic velocity
$\mathbf{u}$ with the pressure gradient $\nabla P(\mathbf{r})$ by the
stationary 2D Euler equations.

For simplicity, we will consider only the pressureless case when equations (%
\ref{3DP}) are written in the form%
\[
\mathbf{u}_{t}+(\mathbf{u\cdot }\nabla )\mathbf{u}+v\mathbf{u}%
_{z}=0,\,\,(\nabla \cdot \mathbf{u})+v_{z}=0.
\]
These equations admit the method of characteristics. However, we will not
consider its application and only notice that the solution given in \cite%
{HongHunter} for the 2D inviscid Prandtl equations can be generalized to the
3D case. From our opinion, the most interesting question is connected with
slipping flow. Its dynamics (at $v|_{z=0}=0$) will be defined by the 2D Hopf
equation
\begin{equation}
\mathbf{u}_{t}+(\mathbf{u\cdot }\nabla )\mathbf{u}=0  \label{Hopf3D}
\end{equation}%
which evidently can also describe breaking of slipping flows. However, the
conditions for breaking in this case are different than in the 2D case (see,
e.g. \cite{konopelchenko2021homogeneous}).

As it was demonstrated in \cite{Kuznetsov2003} solution of equation (\ref%
{Hopf3D}) can be constructed more or less easy if one considers the velocity
gradient $U_{ij}=\partial u_{i}/\partial x_{j}$. Differentiation of equation
(\ref{Hopf3D}) with respect to $\mathbf{r}$ leads to the following matrix
equation
\begin{equation} \label{U}
\frac{dU}{dt}=-U^{2}
\end{equation}
where%
\[
\frac{d}{dt}=\frac{\partial }{\partial t}+(\mathbf{u\cdot }\nabla )
\]
is the (material) derivative along the trajectory:%
\[
\frac{d\mathbf{r}}{dt}=\mathbf{u(r,}t),\,\,\mathbf{r}|_{t=0}\mathbf{=a.}
\]
On this trajectory
\[
\frac{d\mathbf{u}}{dt}=0,\mathbf{u}|_{t=0}\mathbf{=u}_{0}\mathbf{(a)}
\]
that gives the mapping
\begin{equation}
\mathbf{r=a+\mathbf{u}_{0}\mathbf{(a)}}t.  \label{mapping}
\end{equation}%

Then the solution of equation (\ref{U}) has the form \cite{Kuznetsov2003}
\[
U=U_{0}(\mathbf{a})(1+U_{0}(\mathbf{a})t)^{-1}
\]
where $U_{0}(\mathbf{a})$ is the initial values of $U$. Here
\[
\widehat{J}=1+U_{0}(\mathbf{a})t
\]%
is the Jacobi matrix for the mapping (\ref{mapping}).

Expanding the matrix $U_{0}(\mathbf{a})$ through its projectors $P^{(n)}$
yields
\begin{equation}
U=\sum_{k}\frac{\lambda _{k}P^{(k)}}{1+\lambda _{k}t}.  \label{expand}
\end{equation}%
Here the projections $P^{(k)}$, being a matrix function of $\mathbf{a}$, are
expressed through the eigenvectors for the direct ($U_{0}(\mathbf{a})\psi
=\lambda \psi $) and conjugated ($\varphi U_{0}(\mathbf{a})=\varphi \lambda $%
) spectral problems for the matrix $U_{0}(\mathbf{a}),$%
\[
P_{ij}^{(k)}=\psi _{i}^{(k)}\varphi _{j}^{(k)},
\]
where the vectors $\psi ^{(k)}$ and $\varphi ^{(m)}$ with different $k$ and $%
m$ are mutually orthogonal:%
\[
\psi _{i}^{(k)}\varphi _{j}^{(m)}=\delta _{km}.
\]
In terms of the vectors $\psi ^{(k)}$ and $\varphi ^{(m)}$ the eiigen value $%
1+\lambda _{k}t$ of the Jacobi matrix can be written as differential
equations,%
\begin{equation}
\varphi _{i}^{(k)}\frac{\partial x_{j}}{\partial a_{i}}\psi
_{j}^{(k)}=1+\lambda _{k}t  \label{connection}
\end{equation}%
which define connections between the physical $x$-space and the space of
Lagrangiam markers $\mathbf{a}$.

It is easily to see that the Jacobian of the mapping (\ref{mapping}) is
expressed in terms of the eigenvalues as follows
\begin{equation}
J=\det \widehat{J}=\prod\limits_{k}(1+\lambda _{k}t).  \label{Jac}
\end{equation}%
From (\ref{expand}) it is seen that the breaking time is given by the
following expression (compare with \textbf{\cite{Frisch,shandarin1989large}}):
\begin{equation}
t_{0}=\min_{k,a}(-\lambda _{k}^{-1})>0.  \label{min}
\end{equation}%
As it was mentioned in \cite{konopelchenko2021homogeneous}, positive value
of $t_{0}$ imposes a few restrictions on eigen values $\lambda _{k}$. First
of all, in the 2D\ case $\lambda _{1,2}$ as roots of the corresponding
characteristic  equation (quadratic relative to $\lambda $) should be real.
The latter means that the equation discriminant takes real values. Secondly,
among $\lambda _{1,2}$ in the $a$-space there should be found at least one
eigen value which provides positiveness of $t_{0}$.

At $t\rightarrow t_{0}$, as it follows from (\ref{expand}),   $U$ transforms
into the degenerate matrix when the only one term survives corresponding to
minimum (\ref{min}) at some $k$ (which we denote further $1$)  and $\mathbf{%
a=a}_{0},$%
\begin{equation}
U\approx \frac{P^{(n)}}{\tau +\beta _{ij}\Delta a_{i}\Delta a_{j}},
\label{asymp}
\end{equation}%
where $\tau =t_{0}-t$ , $\Delta \mathbf{a}=\mathbf{a}-\mathbf{a}_{0}$,
\[
2\beta _{ij}=-\frac{\partial ^{2}\lambda _{n}^{-1}}{\partial a_{i}\partial
a_{j}}\mid _{a=a_{0}},
\]
and $P^{(n)}$ is a constant projector taken at\ $\mathbf{a=a}_{0}$.
Respectively, at $t=t_{0}$, $k=1$ and $\mathbf{a=a}_{0}$ the Jacobian (\ref%
{Jac}) vanishes.

This gives simultaneous singularities for both symmetric and antisymmetric
parts of $U$. The symmetric part of $U$,
\[
S=\frac{1}{2}(U+U^{T})
\]
has a meaning of the stress tensor where $T$ denotes transpose. The
antisymmetric part of $U$,
\[
\Omega =U-U^{T},
\]
is the vorticity tensor. The presence of the vorticity in the slipping
boundary  means a fluid rotation existence.  Evidently, that while
approaching a singularity the rotation velocity will increase. On the other
hand,  as we saw already for both  2D Prandtl and Euler equations, the
appearance of the breaking/folding for slipping flows leads to the onset of
a jet in the perpendicular direction to the boundary. For 3D Prandtl flows
such jet also exists. The interference of rotation and jet gives an
updraft with rotation \textbf{which is an analogue} of a tornado in this problem.

As in the 2D case (compare with (\ref{1Dsingularity})) for the 3D Prandtl
slipping flow it is possible to find a spatial  structure of a singularity
as $\tau \rightarrow 0$ . Note first that  dependence (\ref{asymp}) \ shows
asymptotically the self similar behavior in the space of Lagrangian markers,
$\Delta a$ $\bowtie $ $\tau ^{1/2}$ in all directions. In the physical $x$%
-space, however, the situation will be very different.  A singularity turns
out to be very anisotropic.   Let $\lambda _{1}$ be an \textbf{eigenvalue} giving
the expansion (\ref{asymp}) then another $\lambda _{2}$ will be finite at
the singular point. In this case in the equation  (\ref{connection})  near
singularity we can consider  $\varphi ^{(1)}$ and $\psi ^{(1)}$ to be
constant vectors (taken at $\mathbf{a=a}_{0}$). Denote as $\Delta
x_{\parallel }$  projection $\Delta \mathbf{x}\cdot \mathbf{\psi }^{(1)}$
and $\Delta a_{\parallel }$ projection $\Delta \mathbf{a}$ corresponding
to direction $\varphi ^{(1)}$. Differential equation for  $x_{\parallel }$,
according to (\ref{connection}), at a small vicinity of the singular point
can be written in the form%
\[
\frac{\partial \Delta x_{\parallel }}{\partial \Delta a_{\parallel }}%
=\lambda _{1}^{-1}\left( \tau +\beta _{ij}\Delta a_{i}\Delta a_{j}\right) ,
\]
where multiplier  $\lambda _{1}^{-1}$ is considered as a constant.  In this
equation the main contribution to the sum $\beta _{ij}\Delta a_{i}\Delta
a_{j}$ near singularity comes from  $\beta _{\parallel }(\Delta a_{\parallel
})^{2}$ where $\beta _{\parallel }$ is easily calculated through components
of the matrix $\beta $ so that asymptotically we have equation coinciding up
to some constants with (\ref{derivative}) for the 1D Hopf equation:
\[
\frac{\partial \Delta x_{\parallel }}{\partial \Delta a_{\parallel }}%
=\lambda _{1}^{-1}\left( \tau +\beta _{\parallel }(\Delta a_{\parallel
})^{2}\right) .
\]
Thus, at $\tau =0$ and $z=0$ the singularity for $U$ has the same structure as for
the 1D Prandtl equation:
\begin{equation}
U\propto \left( \Delta x_{\parallel }\right) ^{-2/3}.  \label{2/3}
\end{equation}%
This dependence is valid almost for all angles except "transverse" one:

\begin{equation}
U\propto \left( \Delta x_{\perp }\right) ^{-2}  \label{2}
\end{equation}%
which  follows from equation (\ref{connection})  for  non-singular $\lambda
_{2}$. Such a dependence on $\Delta x_{\perp }$ means that $\Delta x_{\perp }%
\propto\Delta a_{\perp }$, it appears for almost transverse direction
where two asymptotics (\ref{2/3}) and (\ref{2}) are comparable, i.e. at $\Delta x_{\perp }%
\propto\left( \Delta x_{\parallel }\right) ^{1/3}$.

\section{Conclusion}

In this paper, we have developed a new concept of the formation of big
velocity gradients with the blow-up behavior or with the exponential in time
increase for the slipping flows in incompressible inviscid fluids. These
processes develop as a folding due to compressible character of the slipping
flows. Namely, the boundary itself introduces some element of
compressibility into flow which, from our point of view, can be considered
as a reason of the singularity formation on the boundary. We have
demonstrated for the 2D and 3D inviscid Prandtl equations that singularities
are formed for the velocity gradient on the wall with slipping boundary
conditions. This process is nothing more than breaking phenomenon which is
well known in gas dynamics since the classical works of famous Riemann. In
particular, we have shown that, besides the velocity gradient blow-up on the
boundary for the 2D inviscid Prandtl flow, the vorticity gradient also
becomes singular in a finite time also on the boundary. It turns out that
the similar behavior takes place for slipping flows within the 2D Euler
equation between two parallel plates. In this case, we have numerically
found that maximum of the velocity gradient is developed also on the plate
with exponential increase in time. Simultaneously, the vorticity gradient
has been shown to demonstrate the double exponential growth in time on the
boundary. These results are in correspondence with the proof given by
Kiselev and {\v{S}}ver{\'{a}}k \cite{kiselev2014small} for the Eulerian flow
inside a disk. We showed numerically, that this process can be considered as
a folding with power dependence between the maximum velocity gradient and
its thickness $\ell$: $\max|u_x|\propto \ell^{-2/3}$. It is interesting to
note that such type of regimes with exponential growth were observed in the
3D Euler equations for the vortical pancake structures and within the 2D
Euler equations for quasi-one-dimensional (filamentous) structures in the
form of quasi-shocks of vorticity \cite%
{AgafontsevKuznetsovMailybaevSereshchenko2022}. In the 3D Prandtl equations
we have shown that the slipping flows demonstrate appearance of the blow-up
behavior for both the velocity stress tensor and the vorticity. From our
point of view, the latter problem becomes principally important for
understanding of which role boundaries play in the collapse formation within
the 3D Eulerian flows.

\section{Acknowledgments}

The authors thank B.G. Konopelchenko, P.-L.Sulem, A.A. Kiselev, T.N.
Shilkin, A.M. Gaifullin, G.E. Volovik and V.E. Zakharov for helpful
discussions and A.Ya. Maltsev for valuable remarks.

\end{document}